\documentstyle[aps,prl,twocolumn,epsfig]{revtex}

\def\be{\begin{equation}}
\def\ee{\end{equation}}
\def\bea{\begin{eqnarray}}
\def\eea{\end{eqnarray}}
\def\bma{\begin{mathletters}}
\def\ema{\end{mathletters}}

\def\0{\overline{0}}

\def\q0{\underline{0}}

\def\L{{\cal L}}
\def\M{{\cal M}}

\def\E{{\cal E}}
\def\T{{\cal T}}
\def\O{{\cal O}}

\def\tr{\mbox{ tr }}

\def\bra#1{\langle#1|} 
\def\ket#1{|#1\rangle}

\def\proj#1{\ket{#1}\!\bra{#1}}

\begin{document}

\draft \wideabs{

\title{
On the continuity of asymptotic measures of entanglement
}

\author{G. Vidal}

\address{
Institute for Quantum Information, California Institute of Technology, Pasadena, CA 91125, USA\\
}

\date{\today}
\maketitle


\begin{abstract}

We show that, in any open set of distillable states, all asymptotic entanglement measures $E(\rho)$ are continuous as a function of (a single copy of) $\rho$, even though they quantify the entanglement properties of $\rho^{\otimes N}$ is the large $N$ limit.

\end{abstract}

\pacs{PACS Nos. 03.67.-a, 03.65.Bz}}

One of the first and central results in entanglement theory establishes that, in the large $N$ limit,
$N$ copies of an Einstein--Podolski--Rosen (EPR) state
\be
\ket{\Phi^{+}} \equiv \frac{1}{\sqrt{2}}(\ket{00} + \ket{11})
\label{EPR}
\ee
can be {\em reversibly} converted into $M$ copies of any other entangled pure state $\ket{\psi}$ by means of local operations and classical communication (LOCC) \cite{BBPS}, possibility that we shall symbolically write as \cite{symbol}
\be
\ket{\psi} \rightleftharpoons \ket{\Phi^{+}}^{\otimes N/M}.
\label{trans}
\ee
Subsequently it has also been shown that entangled mixed states $\rho$ exist for which, in contrast, the asymptotic conversion into EPR pairs using LOCC is unavoidably {\em irreversible} \cite{VC},
\be
 \rho \not \rightleftharpoons \ket{\Phi^{+}}^{\otimes N/M}.
\label{transmix}
\ee

Since LOCC can not generate entanglement but only introduce classical correlations, the reversibility of the asymptotic transformation of Eq. (\ref{trans}) indicates that $\ket{\Phi^{+}}^{\otimes N}$ and $\ket{\psi}^{\otimes M}$ contain the same amount of entanglement. Thus for pure states the asymptotic ratio $N/M$, or entropy of entanglement $E(\psi)$ \cite{BBPS}, quantifies the amount of EPR entanglement contained in $\ket{\psi}^{\otimes M}$ per copy of $\ket{\psi}$. 
The irreversibility of the asymptotic transformation of Eq. (\ref{transmix}) implies that, instead, more than one measure is required in order to completely quantify the entanglement of $\rho$ with respect to its asymptotic interconvertibility into EPR pairs using LOCC. In particular, the asymptotic transformations 
\bea
\rho  \rightarrow \ket{\Phi^{+}}^{\otimes N_d/M} \label{distil} \\
\rho  \leftarrow \ket{\Phi^{+}}^{\otimes N_c/M} \label{concen}
\eea
can be characterized by the ratio $N_d/M$, or distillable entanglement $E_d(\rho)$ \cite{BDSW,dist}, where $N_d$ is the maximal number of EPR pairs that can be distilled from $\rho^{\otimes M}$, and also by the ratio $N_c/M$, or entanglement cost $E_c(\rho)$ \cite{BDSW,cost}, where $N_c$ is the minimal number of EPR pairs needed to create $\rho^{\otimes M}$. Since $E_d$ and $E_c$ are inequivalent \cite{VC}, both measures need to be independently considered when quantifying mixed--state entanglement.

Several other measures have been proposed to quantify the entanglement of bipartite mixed states according to their asymptotic convertibilities using LOCC (see \cite{MH} and references therein). In the axiomatic approach any such measure is required to fulfill a series of postulates. In general, it is very difficult to compute any of these measures \cite{AEJPVM}, but it is still possible to study some of their general properties. It has been shown, for instance, that any function $E(\rho)$ compatible with some axioms is confined between the distillable entanglement $E_d$ and the entanglement cost $E_c$ \cite{Ho00},
\be
E_d(\rho) \leq E(\rho) \leq E_c(\rho).
\ee

Here we shall prove another general property of asymptotic measures of entanglement, namely that in any open set $\Gamma$ of distillable states, $E_d >0$, any measure $E$ is continuous as a function of a single copy of $\rho\in \Gamma$, even though $E(\rho)$ actually quantifies the entanglement of $\rho^{\otimes N}$ in the limit $N\rightarrow \infty$. 

We emphasize that the continuity of a function $G(\rho)$ does not imply the continuity of an asymptotic measure $E(\rho)$ that coincides with its regularized version \cite{nocont},
\be
E(\rho) = \lim_{N\rightarrow \infty} \frac{G(\rho^{\otimes N})}{N}, 
\label{dolim}
\ee
and therefore the present result is not implied by the findings of \cite{cont}, where a proof of the continuity of some non-regularized measures is presented. Actually, whether asymptotic measures of entanglement are discontinuous at the border between distillable and non--distillable states remains an open question. We show, however, that all measures are continuous at this border for $2\times 2$ and $2\times 3$ systems. In addition, in $2\times N$ systems the distillable entanglement turns out to continuously vanish at the mentioned border, this being also the case for arbitrary $M \times N$ systems in the catalytic LOCC setting \cite{JP} or when entangled states that have a positive partial transposition (PPT) are available as a resource together with LOCC.

We start by proposing an alternative, simplified approach to the conditions to be required to any asymptotic measure of entanglement. Let
\be
\rho^{\otimes x} \rightarrow \sigma^{\otimes y}, 
\label{asytrans}
\ee
$x,y \geq 0$,
denote a transformation achievable in the asymptotic limit using LOCC \cite{asym}.
Then, the only requirement we impose for $E$ to be an asymptotic measure of entanglement is that \cite{cond}
\be
E(\rho) \geq \frac{y}{x} E(\sigma).
\label{condition}
\ee
In order to be able to compare different measures fulfilling (\ref{condition}), it is convenient to additionally rescale $E$ so that 
\be
E(\ket{\Phi^{+}}) = 1,
\label{norm}
\ee
which corresponds to choosing the entanglement of an EPR state or {\it ebit} as the unit of these measures. 
The main result of this paper is the last of the following series of propositions.

\vspace{2mm}

\noindent {\bf Proposition 1:} $E(\rho^{\otimes N})=NE(\rho)$.

\vspace{2mm}

\noindent{\bf Proposition 2:} $E(\rho)=0$ for any separable state $\rho$.

\vspace{2mm}

\noindent{\bf Proposition 3:} $E(\rho) \geq E(\L(\rho))$ if $\L$ denotes a deterministic LOCC transformation.

\vspace{2mm}

\noindent{\bf Proposition 4:} $E(\rho)$ is the entropy of entanglement for pure states.

\vspace{2mm}

\noindent{\bf Proposition 5:} $E_d(\rho) \geq E(\rho) \geq E_c(\rho)$.

\vspace{2mm}

\noindent{\bf Proposition 6:} $E(\rho)$ is a continuous function in any open set of distillable states.

\vspace{2mm}

All these propositions follow straightforwardly from condition (\ref{condition}) when applied to some specific LOCC transformations, with propositions 4 and 5 assuming also the normalization (\ref{norm}). Indeed, for proposition 1 we need to consider the trivial LOCC transformation
\be
(\rho)^{\otimes Nx} \rightleftharpoons  (\rho^{\otimes N})^{\otimes x}.
\ee
Proposition 2  uses the fact that a separable $\rho_s$ can be reversibly converted into $\rho_s^{\otimes 2}$ using LOCC, so that $E(\rho_s)=2E(\rho_s)$. Proposition 3 follows from the fact that if $\sigma =\L(\rho)$, then for any number of copies $N$, $\rho^{\otimes N}$ can be transformed into $\sigma^{\otimes N}$ by applying $\L$ on each copy of $\rho$. In proposition 4 we need to consider the reversibility of the asymptotic conversion of Eq. (\ref{trans}), whose ratio is given by the entropy of entanglement \cite{BBPS}. Proposition 5 \cite{Ho00} employs the optimal asymptotic distillation strategy using LOCC, Eq. (\ref{distil}), which implies that $E(\rho) \geq E_d(\rho) E(\Phi^+)=E_d(\rho)$, and also the optimal asymptotic preparation protocol using LOCC, Eq. (\ref{concen}) 
which implies that $E(\rho) \leq E_c(\rho)E(\Phi^+) = E_c(\rho)$. 

Finally, we move to prove proposition 6. In order to discuss the continuity of entanglement measures we introduce the trace distance $\T(\rho,\sigma) \equiv \tr |\rho-\sigma|/2$ \cite{book}, which is both a measure of distinguishability and a metric distance on the set of density matrices. Let $\Gamma$ be an open set of distillable states, i.e. $E_d(\rho) > 0$ for all $\rho \in \Gamma$, and consider an arbitrary $\rho \in \Gamma$. Our aim is to show that $E$ is continuous at $\rho$. Let $\epsilon > 0$ be such that the ball $B \equiv \{\sigma ~|~ \T(\rho,\sigma) \leq \epsilon \}$ is a subset of $\Gamma$ (see figure). We consider an arbitrary $\sigma$ on the surface of $B$, i.e. $\T(\rho,\sigma)=\epsilon$, and the one-dimensional family of states 
\be
\rho_p \equiv (1-p)\rho + p\sigma \in B,
\label{family}
\ee
$p \in [-1,1]$.

\begin{figure}
\includegraphics{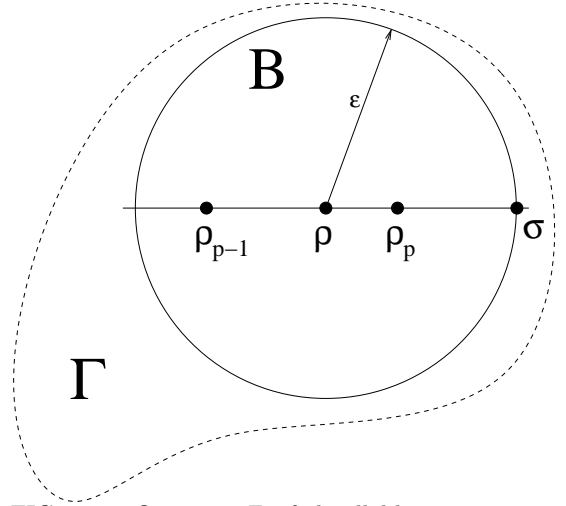}
\caption{\label{fig} Open set $\Gamma$ of distillable states containing a ball $B$ of radius $\epsilon$ and centred in $\rho$. The density matrices $\rho, \rho_{p}, \rho_{p-1}$ and $\sigma$ are related by $\rho_p = (1-p)\rho + p \sigma$ and $\rho = (1-p)\rho_p+p\rho_{p-1}$.}
\end{figure}

Below we will show the feasibility of the asymptotic LOCC transformations
\bea
\rho &\rightarrow& \rho_p^{\otimes (1-\kappa(p))} \label{forth}\\
\rho^{\otimes (1-\kappa(p))} &\leftarrow& \rho_p,
\label{back}
\eea
where $p \in [0,1]$ and $\kappa(p)\equiv p/(p+r/(1-r)) \leq p(1-r)/r$. The constant $r \equiv E_d(B)/E_c(B)$ is defined as the ratio between the minimum of $E_d$ in $B$ and the maximum of $E_c$ in $B$. Then, condition (\ref{condition}) applied to transformations (\ref{forth}-\ref{back}) implies
\bea
E(\rho) &\geq& (1-\kappa(p))E(\rho_p)\\
E(\rho_p) &\geq& (1-\kappa(p))E(\rho),
\eea
or
\be
\kappa(p)E(\rho) \geq E(\rho) - E(\rho_p) \geq - \kappa(p)E(\rho_p).
\ee
Since $E(\rho),E(\rho_p) \leq E_c(B)$, we can write
\be
|E(\rho) - E(\rho_p)| \leq \kappa(\rho)E_c(B) \leq p\Delta,
\ee
where $\Delta \equiv E_d(B)-E_c(B)$.
Notice that $\tr|\rho-\rho_p| = p \tr|\rho-\sigma| = 2p\epsilon$, which allows us to replace the parameter $p$ with $\epsilon$ and $\T(\rho,\rho_p)$. We conclude that, for any $\rho \in \Gamma$ and any $\rho'$ in a ball $B_{\rho,\epsilon} \subset \Gamma$ of radius $\epsilon$ and centred at $\rho$, we have
\be
|E(\rho) - E(\rho')| \leq \frac{\Delta(B_{\rho,\epsilon})}{\epsilon}\T(\rho,\rho'),
\ee
which in particular proves proposition 6. Notice that, significantly, for $\epsilon$ sufficiently small, $\Delta(B_{\rho,\epsilon})$ can be approximated by $E_c(\rho)-E_d(\rho)$, so that the variations of $E$ around $\rho$ are constrained by the gap between $E_c(\rho)$ and $E_d(\rho)$, that is, by the degree of irreversibility in the asymptotic distillation/preparation cycle of Eq. (\ref{transmix}).

Let us then move to justify transformations (\ref{forth})-(\ref{back}). The ingredients we need to use are ($i$) the notions of asymptotic distillability and preparation, which for large $N$ allow e.g. to prepare $N E_d(\rho)/E_c(\sigma)$ copies of $\sigma$ from $N$ the state $\rho^{\otimes N}$,
\be
\rho \rightarrow \sigma^{\otimes E_d(\rho)/E_c(\sigma)},
\label{mixtomix}
\ee 
and ($ii$) the notion of asymptotic mixture, that we next describe.

\vspace{2mm}

\noindent {\bf Lemma:} Asymptotic mixing, represented by the transformation
\be
\rho^{\otimes (1-p)}\otimes\sigma^{\otimes p} ~~\rightarrow~~ (1-p)\rho + p\sigma,
\ee
can be achieved by means of LOCC.

\vspace{2mm}

{\em Proof:} We show how to use $N(1-p)+o(N)$ copies of the state $\rho$ and $Np+o(N)$ copies of the state $\sigma$ to produce a state $\Pi_{\rho_p}^{N}$ which is asymptotically indistinguishable from $N$ copies of the state $\rho_p = (1-p)\rho + p\sigma$, that is $\lim_{N\rightarrow \infty} \T(\rho_p^{\otimes N},\Pi_{\rho_p}^{N})=0$, which proves the lemma. Consider the expansion $\rho_p^{\otimes N}$ as
\be
\rho_p^{\otimes N} = \frac{1}{2^N}\sum_{l=0}^{N} 
\frac{N!}{(N-l)!l!}
S(\rho^{\otimes N-l} \otimes \sigma^{\otimes l}),
\ee
where $S(\rho^{\otimes N-l} \otimes \sigma^{\otimes l})$ denotes the trace--normalized, equally--weighted mixture of all possible permutations of $N-l$ copies of $\rho$ and $l$ copies of $\sigma$. We define
\be
\Pi_{\rho_p}^{N} \equiv \frac{1}{(1-t_N)2^N}\sum_{l=Np-N^{2/3}}^{Np+N^{2/3}} 
\frac{N!}{(N-l)!l!}
S(\rho^{\otimes N-l} \otimes \sigma^{\otimes l}),
\label{Pi}
\ee
where $t_N$ is the (asymptotically vanishing) trace of the discarded tails of the binomial distribution. Notice that at most $N(1-p)+N^{2/3}$ copies of $\rho$ and $Np+N^{2/3}$ copies of $\sigma$ are required to prepare any of the states that appear in the mixture (\ref{Pi}), and that classical communication (or just pre--shared randomness) suffices for two distant parties to be able to perform the mixture. Finally, it is easy to derive the upper bound $|\rho_p^{\otimes N}-\Pi_{\rho_p}^{N}| \leq 2t_N$, which finishes the proof. $\Box$

Let us consider first transformation (\ref{forth}). Starting from $N$ copies of $\rho$, a fraction of $\delta N$ copies can be asymptotically transformed into $\delta N r$ copies of $\sigma$ (cf. \ref{mixtomix}), where $r=E_d(B)/E_c(B)\leq E_d(\rho)/E_c(\sigma)$, 
\be
\rho \rightarrow \rho^{\otimes 1-\delta} \otimes \sigma^{\otimes \delta r}.
\ee 
Then, by choosing $\delta = p/(r+(1-r)p)$ and performing asymptotic mixing as indicated above, $N(1-\kappa(p))$ copies of $\rho_p$ are obtained. Transformation (\ref{back}) can be obtained using the same method, but starting from $N$ copies of $\rho_p$ and converting a fraction of $\delta N$ copies into $\delta N r$ copies of $\rho_{p-1}$ (cf. Eq. (\ref{family}) and figure), so that the product of the mixing are $N(1-\kappa(r))$ copies of $\rho$.  This completes the proof of the main result of this paper.

We have therefore seen that asymptotic measures of entanglement, defined with respect to LOCC transformations, are continuous in any open set of distillable states. 
This result does not exclude that measures have a discontinuity ($i$) at the surface of the set of density operators, in particular when moving from mixed states to pure states, and ($ii$) at the border between distillable and non--distillable states. 

However, known LOCC protocols can be again used to prove the continuity of asymptotic entanglement measures in some particular cases. For instance, a measure $E$ fulfilling condition (\ref{condition}) is continuous at the point specified by $\rho=\proj{\Phi^+}$. In order to see this we need to consider the distillation protocol discussed in \cite{BDSW}, that for sufficiently small $\epsilon > 0$, allows to asymptotically transform copies of $(1-\epsilon)\proj{\Phi^+} + \epsilon \xi$, where $\xi$ represents an arbitrary state, into copies of $\ket{\Phi^+}$ with a ratio $1-\eta(\epsilon)$ that goes continuously to 1 as $\epsilon$ decreases, 
\be
(1-\epsilon)\proj{\Phi^+} + \epsilon \xi ~~\rightarrow ~~\proj{\Phi^{+}}^{\otimes 1- \eta(\epsilon)}.
\ee
Since this transformation can be inverted with at least unit ratio, applying condition (\ref{condition}) implies the continuity of $E$ also at $\ket{\Psi^+}$ \cite{altres}. 
Another example is the continuity of all measures at the border between distillable and non--distillable entangled states of $2 \times 2$ and $2 \times 3$ systems. In these cases the distillable entanglement $E_d$ ---and therefore also the rest of measures--- is non-zero for all inseparable states \cite{HW}. We only need to show that the entanglement cost $E_c$ (an upper bound to the rest of measures) continuously goes to zero when approaching the set of separable states from the set of distillable ones. But this is the case, since $E_c(\rho) \leq E_f(\rho)$, where the entanglement of formation $E_f(\rho)$ is a continuous function \cite{cont} that only vanishes for separable states. More generally, we can also argue that $E_d$ is continuous in the border between distillable and non--distillable states in a $2\times N$ system, since in this case states are distillable if and only if they have a non--positive partial transposition NPPT \cite{HW}, and the logarithmic negativity \cite{neg} is a continuous upper bound to $E_d$ that vanishes at the border. However, we do not know whether entanglement measures are discontinuous at the border of distillable states for systems other than $2\times 2$ and $2\times 3$.

Finally, one can also define asymptotic measures of entanglement in relation to other classes of allowed operations. Given a class $C$ of operations, we again require condition (\ref{condition}) to a measure $E^C(\rho)$ whenever transformation (\ref{asytrans}) is achievable by operations in class $C$.

For instance, in addition to LOCC operations, entangled states with positive partial transposition (PPT) can be made available as an extra resource. In this case all NPPT states can be distilled \cite{distppt}, and the upper bound to distillable entanglement given by the logarithmic negativity \cite{neg} still holds. This implies that $E_d^{LOCC+PPT}$ continuously vanishes at the border between PPT and NPPT states. 

Catalytic LOCC operations \cite{JP} are another popular class of operations. Let us consider the following chain of asymptotic transformations:
\bea
 \rho \otimes \proj{\Psi^+}^{\otimes \delta}\rightarrow\\
 \rho \otimes \sigma^{\otimes \delta/E_c(\sigma)}\rightarrow\\
 \rho_p^{\otimes 1+\delta/E_c(\sigma)} \rightarrow\\
 \rho_p^{\otimes 1+\delta/E_c(\sigma)-\delta/E^{cat}_d(\rho_p)} \otimes \proj{\Psi^+}^{\otimes \delta},
\eea
where $p=\delta/E_c(\sigma)/(1+\delta/E_c(\sigma))$ and we have used, respectively, the asymptotic processes of preparation, mixing and (catalytic) distillation. That is, by borrowing sufficiently many copies of the pure state $\ket{\Psi^+}$, $N$ copies of $\rho$ can be asymptotically transformed into $N( 1+\delta/E_c(\sigma)-\delta/E^{cat}_d(\rho_p))$ copies of $\rho_p = (1+p) \rho +p \sigma$, provided $E_d(\rho_p) > 0$. This means, applying (\ref{condition}), that any measure $E^{cat}$ in the catalytic LOCC setting fulfills
\be
E^{cat}(\rho) \geq (1+\delta (\frac{1}{E_c(\sigma)} - \frac{1}{E^{cat}_d(\rho_p)}))E^{cat}(\rho_p).
\ee
We can now fix $\sigma$ and $\rho_p$ so that $E_c(\sigma), E_d(\rho_p)>0$ and that $\rho$ depends on $\delta$ (or $p(\delta)$). In this case we have
\be
E^{cat}(\rho) \geq (1+\delta k)E^{cat}(\rho_p),
\label{cat}
\ee
where $k\equiv 1/E_c(\sigma) - 1/E_d(\rho_p)$, $|k| < \infty$. Notice that $\rho$ continuously approaches $\rho_p$ as $\delta$ decreases, $\T(\rho,\rho_p) = p\T(\rho,\sigma)$. This can be used to show, for instance, that Eq. (\ref{cat}) for $E^{cat}=E_d^{cat}$ implies the continuity of $E_d^{cat}$ at the border between states that can be distilled using catalytic LOCC and those that can not.

The author thanks J. Ignacio Cirac, David P. DiVincenzo, Patrick Hayden, Alexei Kitaev, Barbara Kraus, Benjamin Schumacher, Barbara M. Terhal and Reinhard F. Werner for useful discussions.
This work was supported by the National Science Foundation (of the United States of America) under Grant. No. EIA-0086038.


\begin{references}
\bibitem{symbol} The formal expression $\sigma^{\otimes x}$ for a real $x\geq 0$ takes meaning in the asymptotic limit, where it is used to denote the state $\rho^{\otimes \lfloor xL \rfloor}$ for large $L$. $\lfloor y \rfloor$ is the greatest integer $M$ such that $y \geq M$.
\bibitem{nocont} A function $E(\rho)$ is continuous at $\rho$ if for any other density matrix $\sigma$ we have $\lim_{\epsilon \rightarrow 0^+} E(\rho+\epsilon\delta) = E(\rho)$, where $\delta \equiv \sigma-\rho$. Notice that for $E(\rho)$ in Eq. (\ref{dolim}) to be continuous we would need
\be
\lim_{\epsilon \rightarrow 0^{+}} \lim_{N \rightarrow \infty} \frac{G([\rho+\epsilon\delta]^{\otimes N})}{N} = \lim_{N\rightarrow \infty} \frac{G(\rho^{\otimes N})}{N}, 
\ee
which does not follow from the continuity of $G$, since for any $\epsilon >0$ the states $\rho^{\otimes N}$ and $[\rho+\epsilon\delta]^{\otimes N}$ become orthogonal as $N$ grows, $\lim_{N\rightarrow \infty}\T(\rho^{\otimes N},[\rho+\epsilon\delta]^{\otimes N})=1$. Instead, we would need an asymptotic behaviour of $G$ in $N$ such that 
\be
|G([\rho+\epsilon\delta]^{\otimes N}) - G(\rho^{\otimes N})| \leq  f(\epsilon\delta) N,
\ee
for some function $f(\epsilon\delta)$ that vanishes in the limit $\epsilon \rightarrow 0^{+}$.
\bibitem{asym} The asymptotic transformation of Eq. (\ref{asytrans}) means, by definition, that a sequence of LOCC transformations $\{\E_N\}$ exists such that they transform the state $\rho^{\otimes \lfloor xN \rfloor}$ into the state $\Pi_\sigma^{\lfloor (y-\delta)N \rfloor}$, where for any $\epsilon,\delta >0$ an integer $N_{\epsilon,\delta}$ exists such that for all $N \geq N_{\epsilon,\delta}$,
\be
\T(\sigma^{\otimes \lfloor (y-\delta)N \rfloor},\Pi_\sigma^{(y-\delta)N}) \leq \epsilon.
\ee
\bibitem{cond} We extend the domain of an asymptotic measure $E(\rho)$ so that it takes values for $\rho_1^{\otimes x_1}\otimes\rho_2^{\otimes x_2}\otimes \cdots \otimes \rho_l^{\otimes x_l}$, $x_1,x_2,\cdots,x_l \geq 0$. Then, for instance, if the asymptotic transformation
\be
\rho_1^{\otimes x_1} \otimes \rho_2^{\otimes x_2} \rightarrow \sigma_1^{\otimes y_1} \otimes \sigma_2^{\otimes y_2},
\label{trans4}
\ee 
$x_1,x_2,y_1,y_2 \geq 0$, is possible using only LOCC, we require 
\bea
&&E(\rho_1^{\otimes \frac{x_1}{x_1+x_2}}\otimes \rho_2^{\otimes \frac{x_2}{x_1+x_2}})\,\, \geq \nonumber \\
&&\frac{y_1+y_2}{x_1+x_2}E(\sigma_1^{\otimes \frac{y_1}{y_1+y_2}}\otimes \sigma_2^{\otimes \frac{y_2}{y_1+y_2}}).
\eea
Notice that, arguing as in proposition 1, we obtain $E(\rho^{\otimes x}) = xE(\rho)$, also 
\be
E(\rho_1^{\otimes yx_1}\otimes \rho_2^{\otimes yx_2}) = yE(\rho_1^{\otimes x_1}\otimes \rho_2^{\otimes x_2}).
\ee
\bibitem{altres} Stochastic LOCC transformations that succeed with great probability for single copies of entangled states are a powerful tool to study the continuity of asymptotic entanglement measures in the multipartite setting. Consider, for instance, an entanglement measure $E$ for multipartite pure states and a two--outcome measurement $\M_\epsilon = (M_1, M_2)$, with measurement operators $M_1 \equiv \sqrt{I-\epsilon A^{\dagger}A}$ and $M_2 \equiv \sqrt{\epsilon} A$, $M_1^{\dagger}M_1+M_2^{\dagger}M_2=I$, where $A$ is an arbitrary operator normalized to $\tr |A^{\dagger}A| =1$,  and $\epsilon>0$ is a small parameter, $\epsilon << 1$. This POVM applied to one subsystem transforms $\ket{\psi}$ into $\ket{\psi_\epsilon} \equiv M_1\ket{\psi}/\sqrt{p} = \ket{\psi} + \O(\epsilon)$ with probability $p= 1 - \O(\epsilon)$, which means that, for large $N$, $N$ copies of $\ket{\psi}$ can be converted into about $N(1-\O(\epsilon))$ copies of $\ket{\psi_\epsilon}$. Because $M_1$ can be inverted, the inverse conversion is also possible with similar ratio, and we conclude, in view of condition (\ref{condition}), that $|E(\psi)-E(\psi_{\epsilon})| \leq \O(\epsilon)$.
\end{references}
\end{document}